\documentclass[useAMS,usenatbib, referee, usegraphicx, mathtools]{biom}
\usepackage{pdfpages}  
\usepackage{amsmath}
\usepackage{outlines}

\def\bSig\mathbf{\Sigma}

\usepackage{adjustbox}
\usepackage[figuresright]{rotating}
\usepackage{multirow}
\usepackage{makecell}
\usepackage{url}
\usepackage{tabularx}
\usepackage{natbib}

\title[A Pseudo-Value Regression Approach for Differential Network Analysis of Co-Expression Data]{A Pseudo-Value Regression Approach for Differential Network Analysis of Co-Expression Data}

\author{Seungjun Ahn$^{1}$, 
Tyler Grimes$^{2}$,and
Somnath Datta$^{1}$ \\
$^{1}$Department of Biostatistics, University of Florida, Gainesville, Florida, U.S.A. \\
$^{2}$Department of Mathematics and Statistics, University of North Florida, Jacksonville, Florida, U.S.A.}

\begin{document}




\pagerange{\pageref{firstpage}--\pageref{lastpage}} 
\pubyear{2022}
\artmonth{October}




\label{firstpage}


\begin{abstract}
The differential network (DN) analysis detects changes in measures of association among genes under two or more conditions. In this study, we introduce a Pseudo-value Regression Approach for Network Analysis (PRANA). This is a novel method of DN analysis that also adjusts for additional clinical covariates. We start from mutual information (MI) criteria, followed by pseudo-value calculations, which are then entered into a robust regression model. Performance in terms of precision, recall, and F1 score of differentially connected (DC) genes is assessed in both univariable and multivariable settings through variety of simulations. By and large, PRANA outperformed \textit{dnapath} and DINGO, neither of which is equipped to adjust for available covariates such as patient-age. Lastly, we employ PRANA in a real data application from the Gene Expression Omnibus (GEO) database to identify DC genes that are associated with chronic obstructive pulmonary disease (COPD) to demonstrate its utility. This is the first attempt of utilizing a regression modeling for DN analysis by collective gene expression levels between two or more groups with the inclusion of additional covariates.
\hfill\\
\end{abstract}
%



\maketitle


%

\section{Introduction}
\label{s:intro}
The rapid advancement of RNA-sequencing (RNA-Seq) data from high-throughput sequencing technologies has provided clear advantages in gene expression studies. It has broadened our understanding of genetics and pathogenesis of human diseases \citep{highthrough1, highthrough2}. Compared to microarrays, RNA-Seq has a wider dynamic range, the ability to detect novel transcripts, and often results in higher sensitivity and specificity of detection of differential gene expression \citep{ref1review2comment1, ref2review2comment1}. With the increase of gene expression studies, the statistical methods to analyze these gene expression data have also accordingly adapted and progressed. The regression modelling for RNA-Seq differential expression (DE) analysis has been established to compare the number of DE genes under different biological or clinical states, including linear model based \textit{limma} \citep{limma}, negative binomial model based \textit{edgeR} \citep{edgeR}, or Poisson log-linear model approach \citep{witten}. The analysis of DE has a major limitation in that it looks at one gene at a time, even though a set of genes are often involved in the same biological process \citep{DElack1, DElack2}. In contrast, the differential network (DN) analysis complements the DE analysis \citep{fuente} by looking at genes collectively. 

The DN analysis identifies changes in measures of association (\textit{i.e.} network properties or topologies) of the networks across biological conditions, which makes it distinct from a single network analysis. Several groups have proposed statistical methods for DN analysis \citep{DINGO, DGCA, Gill, dnapath}. In particular, DINGO \citep{DINGO} and \textit{dnapath} \citep{dnapath} have developed methods for RNA-Seq data, to find differentially connected (DC) genes in subnetworks corresponding to different pathways, between two groups of patients; \textit{e.g.} `high-risk' vs. `low-risk' or `long-term survivors' vs. `short-term survivors.' While these methods are convenient to use and applicable, they do not consider other observed covariates that may be associated with gene expression.

For instance, a previous study has shown that the expression levels of oxidative stress-associated genes were differentially expressed with smokers with chronic obstructive pulmonary disease (COPD) through gene set enrichment analysis using microarray data \citep{clin1}. Let us suppose we want to carry out DN analysis on expression data that includes oxidative stress-associated genes and smoking status, which would be used as a grouping variable. In practice, clinicians would also want to include additional covariates such as patient history of cardiac arrhythmia \citep{clin2} and lung carcinoma \citep{clin3} to garner more information for better prognosis. However, there is no available direct regression modeling for DN analysis regressing gene expression level between the smoking statuses with the inclusion of additional clinical covariates described above.

The pseudo-value approach was first developed from the leave-one-out jackknife subsampling procedure, applied to a marginal quantity representing some aspect of a marginal distribution of the response variable. It was originally introduced by Andersen and his colleagues \citep{pseudoref1, pseudoref2} for multi-state survival models. Several studies \citep{pseudoref3, pseudoref4} purported that the pseudo-value regression has advantages that the pseudo-values derived from an asymptotically linear and unbiased estimator are approximately independent and identically distributed with the same conditional expectation. Ahn and Logan \citep{pseudoref5} and Ahn and Mendolia \citep{pseudoref6} showed that their pseudo-value approaches controlled the type I error while maintaining high power with clustered survival data. With these benefits, we propose a regression modeling method that regresses the jackknife pseudo-values \citep{pseudo.efron} derived from a measure of connectivity of genes in a network to estimate the effects of predictors. Note that the grouping variable itself could also be included in the regression model along with additional clinical covariates while regressing the pseudo-values. We loosely refer to this as a ``multivariate setting'', whereas in ``univariate settings'' only the group variable is utilized in a DN analysis.

Thus, in this paper, we introduce a Pseudo-value Regression Approach for Network Analysis (PRANA). This is a novel method of DN analysis that can adjust for additional covariates. We start from mutual information (MI) criteria, followed by pseudo-value calculations, which are then entered into a robust regression model. This article assesses the model performances of our pseudo-value approach in a multivariable setting, followed by a comparison to \textit{dnapath} and DINGO in the univariable setting through simulations. Lastly, we employ our method in a real data application \citep{realdata} from the Gene Expression Omnibus (GEO) database \citep{GEOdatabase} to identify DC genes that are associated with COPD. All statistical analyses are performed in \textsf{R} version 4.0.2 (\textsf{R} Foundation for Statistical Computing, Vienna, Austria).

\section{Methods}
\label{s:model}

\subsection{Mutual Information and ARACNE}
Mutual information (MI) determines whether and how two genes interact. That is, it is a measure of their relatedness and calculated from their joint expression profiles. MI is zero if and only if the joint distribution between the expression level of gene $j$ and gene $k$ satisfies $P(g_j, g_k) = P(g_j)P(g_k)$ for $j \neq k$, or if $j = k$; in other words, they are statistically independent. Algorithm for the Reconstruction of Accurate Cellular Networks (ARACNE) which are proposed by \citet{ARACNE} and \citet{timedelayARACNE} estimates MI using a computationally efficient Gaussian kernel estimator. The estimate of MI is used to quantify the connectivity of each pair of genes in a network. Given a set of two genes measurements, $\overrightarrow{u_i} \equiv ( g_{ij}, g_{ik} )$, $i = 1, \ldots, n$, the joint probability distribution is approximated as $f(\overrightarrow{u}) = 1/n \sum_i h^{-2} \Phi (h^{-1} (\overrightarrow{u} - \overrightarrow{u_i}))$, where $\Phi$ is the bivariate standard normal density and $h$ is the position-dependent kernel width. Then the MI can be expressed as \citep{ARACNE}:
\[
\hat{I}_{jk} = \frac{1}{n} \sum_i \log \frac{f(g_{ij}, g_{ik})}{f(g_{ij}) f(g_{ik})},
\]
where $f(g_j)$ and $f(g_k)$ are the marginals of $f(\overrightarrow{u})$. The matrix containing entries $\hat{I}_{jk}$ is defined as the association matrix.
The ARACNE algorithm copula-transform the profiles for MI estimation because MI is reparameterization invariant; thus, the range of these transformed variables is between 0 and 1 \citep{zeroone}.

\subsection{Pseudo-Value Approach}

Let $\hat{I}_{jk}$ be the MI estimate for a pair of genes $j,k \in \{1, \dots , p\}$ of an estimated network from $n$ individuals. For each gene $k$, we sum the edges (MI estimates) around the gene by taking the column sum of the association matrix to obtain the total connectivity (which can be deemed as a continuous version of degree centrality) of gene $k$:
\begin{equation*}
    \hat{\theta}_{k} = \sum_{j=1}^{p} \hat{I}_{jk},
\end{equation*}
where $k = 1, \dots , p$. 

The jackknife pseudo-values \citep{pseudo.efron} for the $i^\text{th}$ individual and $k^\text{th}$ gene are defined by:
\begin{equation}
\tilde{\theta}_{ik} = n\hat{\theta}_{k} - (n-1)\hat{\theta}_{k(i)}, \label{eq:2.1}
\end{equation}
where $\hat{\theta}_{k(i)}$ is the column sum of a gene calculated from the re-estimated association matrix using the RNA-Seq data without the $i^\text{th}$ subject. For each gene $k$, the re-estimation process requires $n$ such calculations with the data size of $n-1$. 

Let $Z$ a binary group indicator. Let $\mathcal{G}_1 = \{i: Z_{i} = 1 \}$, $\mathcal{G}_2 = \{i: Z_{i} = 2 \}$, and $n_z = |\mathcal{G}_z|$ is the sample size for the two groups $z=1, 2$ and $n = \sum n_{z}$. The jackknife pseudo-values are separately obtained within groups. Following the general formula above, for gene $k$ and group $z$, we similarly define $\hat{\theta}_{k}^z$ and $\hat{\theta}_{k(i)}^z$, where $i = 1,\ldots, n_{z}$. Then for each $i \in \mathcal{G}_z$, the $k^\text{th}$ gene jackknife pseudo-values are calculated by $\tilde{\theta}_{ik} = n_z\hat{\theta}_{k}^z - (n_z-1)\hat{\theta}_{k(i)}^z$.

Next, a robust regression is applied to regress the pseudo-values on a set of covariates, including $Z$ and $\textbf{X}$, where $Z$ is the group indicator and $\textbf{X} = (X_{1}, \dots , X_{q}$) are the potential confounders, such as age and gender. 
For the $i^\text{th}$ individual and $k^\text{th}$ gene, we posit the model
\begin{equation}
E[\tilde{\theta}_{ik} | Z_{i}, \textbf{X}_{i}] = \alpha_{k} + \beta_{k}Z_{i} +  \gamma_{k1}X_{i1} + \dots + \gamma_{kq}X_{iq}, \label{eq:2.2}
\end{equation} 
where $\alpha_{k}$ is the intercept, $\beta_{k}$ is the regression coefficient for $Z$, and $\gamma_{k1}, \dots , \gamma_{kq}$ is the set of regression coefficients to be estimated for $X$. The main parameter of interest is $\beta_{k}$ to test for the change in total connectivity (or degree centrality) of the $k^{\text{th}}$ gene between groups.
Least trimmed squares (LTS), also known as least trimmed sum of squares \citep{LTS}, is then implemented to perform a robust regression. The LTS estimator is defined by
\[
\min_{\alpha_k, \beta_k, \gamma_{k1}, \ldots, \gamma_{kq}} \sum_{i=1}^h r_{(i)} (\alpha_k, \beta_k, \gamma_{k1}, \ldots, \gamma_{kq})^2,
\]
where $r_{(i)}$ is the set of ordered absolute values of the residuals (in increasing order of absolute value) and $h$ may depend on some pre-defined trimming proportion $c$, for instance by means of $h = [n(1-c)] + 1$. In general, $c$ is chosen between 0.5 and 1 \citep{metrika}.

\subsection{Hypothesis Testing}
To test whether the true difference in total connectivity of $k^\text{th}$ gene differs between groups, we test the null hypothesis of $H_{0}: \beta_{k} = 0$ against the research hypothesis $H_{1}: \beta_{k} \neq 0$. The \textit{t}-statistic is computed by $\hat{\beta}_{k}/SE(\hat{\beta}_{k})$, where $SE(\hat{\beta}_{k})$ standard error of $\hat{\beta}_{k}$, obtained using large sample theory, and which is the least-squares estimator of $\beta_{k}$ for $k = 1, \dots , p$ from the robust regression described in equation (\ref{eq:2.2}). P-values are calculated using a \textit{t}-distribution as in \textit{robustbase} \textsf{R} package \citep{robustbase1}.

It is important to control the false discovery rate (FDR), since multiple hypothesis tests are conducted in the DN analysis. The FDR measures the proportion of false discoveries among a set of genes which are significantly DC between groups. The empirical Bayes screening (EBS) approach \citep{eBayes2005} has been applied to control the FDR, which is an extension of Westfall and Young step-down adjusted p-values \citep{WY}. The EBS procedure is robust against model mis-specification, as it utilizes nonparametric function estimation techniques for the estimation of the marginal density of the transformed p-values.

\section{Materials}
This section details step-by-step procedures how the simulation is performed. The performance of our proposed method is assessed by an extensive simulation study. Data are simulated with different number of genes $p$ and sample size $n$. In this simulation, the regression model includes two covariates $Z$ and $X$, where $Z$ is the group indicator and $X \sim N(55, 10)$ is a continuous covariate such as the age of a patient. Three different simulation scenarios are considered.

\subsection{Data Generation}

Simulate weighted networks and RNA-Seq data with a dependence structure that depends on $Z$ and/or $X$ using the \textit{SeqNet} R package \citep{seqnet}. In this setting, there are total of six networks for the combination of two groups and three age categories (younger than 50, 50-60, and older than 60). We consider three different scenarios incorporating group information only (scenario I), age and group information (scenario II), and age and group information with unequal sampling proportions with different distributions of the age in the two groups (scenario III) (see Figures 1--3 for visual demonstrations). \medskip

\subsubsection*{Scenarios I (a--b) and II (a--c)}

\begin{itemize}
    \item[a.] Generate the first random network with $p$ nodes for $z = 1$. The $p \times p$ adjacency matrix, where the diagonal elements are 0 and non-diagonal elements are in $\{0, 1\}$, is extracted from this first graph. It is a symmetric matrix indicating whether a pair of nodes are connected by an edge. Take the column sum of the adjacency matrix to see the total number of connected edges to the node. Record the indices of this vector with column sum for the use of effect size adjustment in later step.

    \item[b.] Perturb the first network to generate the second network for $z = 2$ by removing the edges around nodes using the indices obtained in previous step. To assess the effect size of group, the top $5\%$, $10\%$, and $20\%$ of total nodes with the most number of edges in a network lose their edges (\textit{e.g.} 2 nodes with the most number of edges for a network with $p = 20$ for the effect size of $10\%$, Figure 1). This is the end of scenario I.

    \item[c.] For scenario II, further perturb remaining networks by removing edges of one additional node with the next most number of edges, coming after Step (b) above. This is to simulate networks with a covariate dependence structure on both age and group (see Figure 2).
\end{itemize}

\subsubsection*{Scenario III}

We created a scenario where age is acting like a confounder. In other words, for a given each category that two networks are the same, but the distributions of the age of the patients are different in the two groups. Therefore, there will be an observed difference in network connectivity, which is explained through age.
\begin{itemize}
    \item[a.] Generate the first random network with $p$ nodes for younger than 50 category. The $p \times p$ adjacency matrix, where the diagonal elements are 0 and non-diagonal elements are either $\{0, 1\}$, is extracted from this first graph. Record the indices of connected edges for the perturbation of network in later steps below.

    \item[b.] Perturb the first network to generate the second network for age 50-60 category by removing the edges of the two nodes with the most number of edges in a network lose all of their edges. In other words, we refer to the indices, recorded in the adjacency matrix from the earlier step, and remove all the connected edges around the two nodes.

    \item[c.] Next, we repeat the same to perturb the second network to obtain the third network for older than 60 category (see Figure 3).
\end{itemize}

\begin{figure}
 \centerline{\includegraphics[scale=0.4]{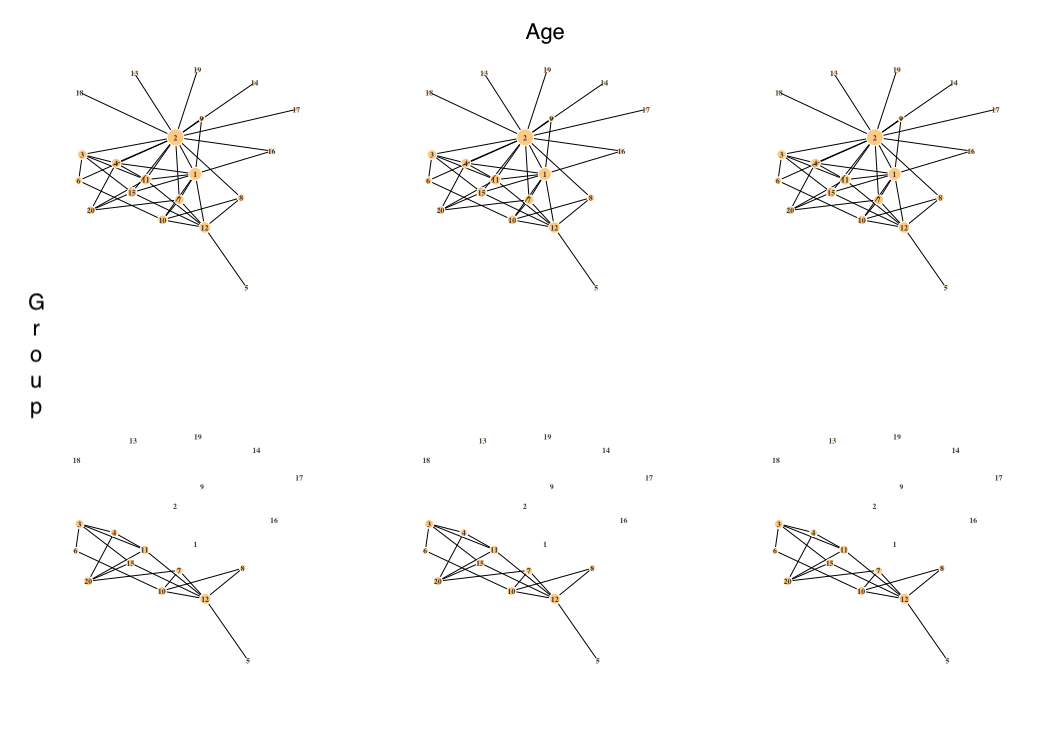}}
  \caption{Network plots visualizing the gene network (\textit{p} = 20) without a covariate dependence structure that depends on binary group only (scenario I). The row represents group whereas the column represents age categories. The three networks in each row are identical, since there is no effect of age on the structure of network. The edges of the hub nodes are removed based on the effect size of the binary group.}
\end{figure}

\begin{figure}
 \centerline{\includegraphics[scale=0.4]{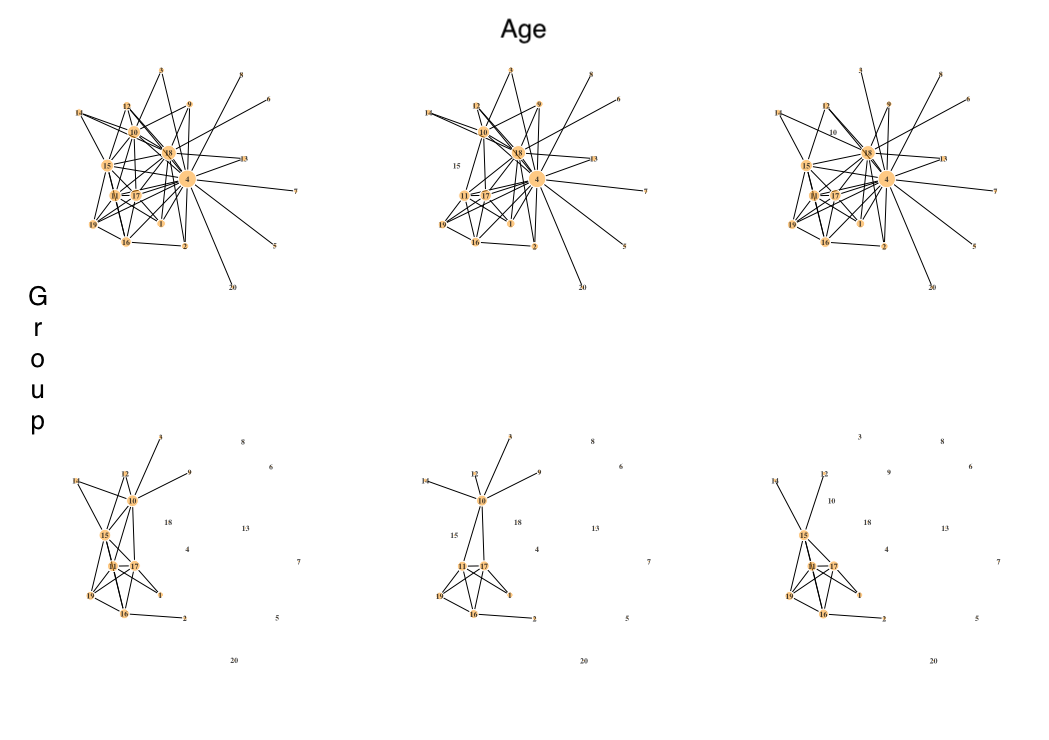}}
  \caption{Network plots visualizing the gene network (\textit{p} = 20) with a covariate dependence structure that depends on age and group information (scenario II). The row represents group whereas the column represents age categories. All six networks have unique structure of the network. The edges of the hub nodes are firstly removed based on the effect size of the group, as shown in Figure 1 above. For this scenario II, additional edges of nodes with greater number of connected edges are removed for each age category.}
\end{figure}

\begin{figure}
 \centerline{\includegraphics[scale=0.4]{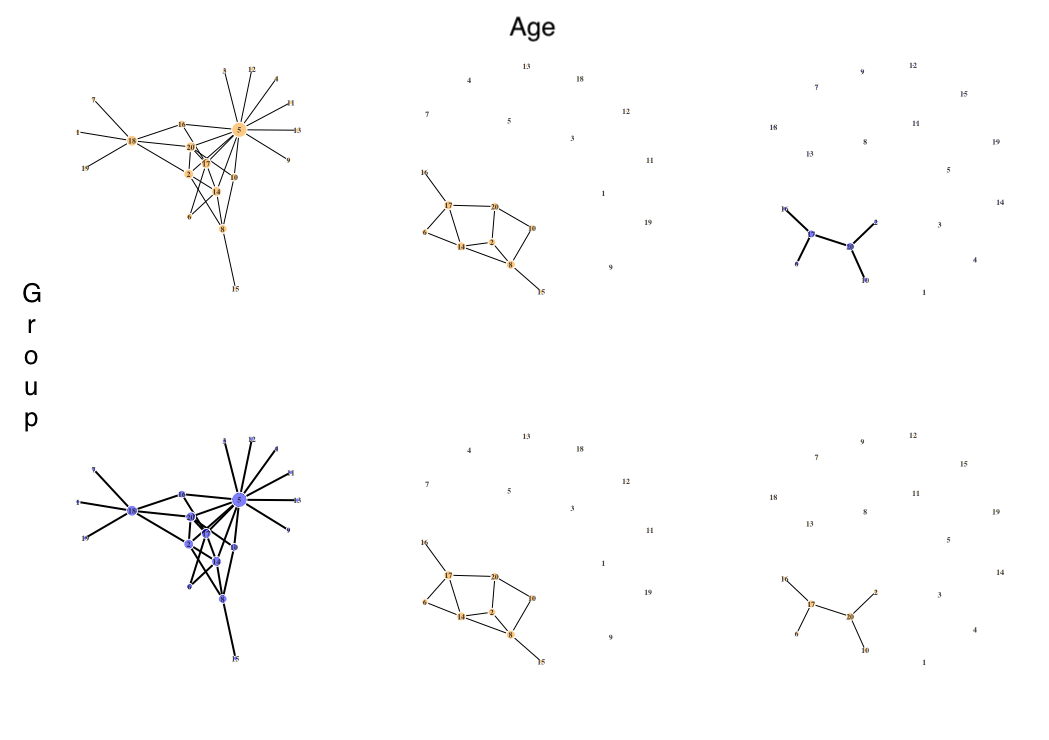}}
  \caption{Network plots visualizing the gene network (\textit{p} = 20) with a covariate dependence structure that depends on age and group information with unequal sampling proportions with respect to different distribution of the age in the two groups (scenario III). The row represents group whereas the column represents age categories. All six networks have unique structure of the network. The edges of the two hub nodes are removed for each age category. To employ the effect of group, $10\%/10\%/80\%$ of the subjects in $z = 1$ will have a network structure to each of the first, second, and third networks in the first row. In contrast, $80\%/10\%/10\%$ of the subjects in $z = 2$ will have a network structure to each of the first, second, and third networks in the second row.}
\end{figure}

\bigskip
For all three scenarios, we assign a partial correlation to edges to obtain weighted networks \citep{seqnet}. Note that adjacency matrices of these weighted networks are used for the the true connection per gene. Generate RNA-Seq samples based on weighted networks with equal sampling proportions for scenarios I and II. However, specifically for scenario III, a sampling proportion differs across age categories and groups. That is, $10\%/10\%/80\%$ for $z=1$ and $80\%/10\%/10\%$ for $z = 2$. The data generation involves with two major steps. Firstly, we generate gene expressions (Gaussian values) from a group-specific weighted network for each gene, denoted as $\Tilde{x}_{i} \sim N(0, 1)$. These Gaussian values are then mapped into RNA-Seq data column-wise by using the inverse CDF of empirical distribution of the reference data using expression data with accession number GSE158699 \citep{realdata} from the Gene Expression Omnibus (GEO) database \citep{GEOdatabase}. We will have $n_{z} \times p$ matrices for each group $z = 1, 2$. 

\subsection{Algorithm}
\begin{outline}[enumerate]
\1 Obtain an association matrix with ARACNE from the data generated in steps from the ``Data Generation'' section to fit an estimated network using \textit{minet} \citep{ARACNE, minet} for each group.
\1 For each gene $k$, calculate the column sums of association matrix for each group $z$ separately, denoted by $\hat{\theta}_{k}^{z}$.
\1 For each gene $k$ and individual $i \in \mathcal{G}_{z}$, compute $\hat{\theta}_{k(i)}^{z}$ from the association matrix that is re-estimated based on RNA-Seq data without the $i^{\text{th}}$ subject of $n_{z} \times p$ data from the ``Data Generation'' section for each group $z$ separately, where $i = 1, \dots , n_{z}$. 

\1 Calculate $\tilde{\theta}_{ik}$ using equation (\ref{eq:2.1}) based on Step 2 and 3. 
\1 For each gene $k$, fit a multivariable robust regression with binary group variable and continuous age variable to obtain the p-values of the group variable, computed from the \textit{t}-test. These p-values are used to compute the performance measures of simulation study. More details on the performance measures are stated next.
\end{outline}

\subsection{Performance Measures}
To evaluate the performance of our proposed method, precision, recall, and the F1 score are calculated. Let $\Omega^{z} \in \mathbf{R}^{p \times p}$ be the adjacency matrix for group $z$, where 
\begin{equation*}
    \Omega_{jk}^{z} =
        \begin{cases}
          1 & \text{if } j^{\text{th}} \text{ gene and } k^{\text{th}} \text{ gene are connected}\\
          0 & \text{otherwise},
        \end{cases}
\end{equation*}
for $z = 1, 2$. Then, for each gene $k$, we calculate
\begin{equation*}
    \eta_{k} = I\bigg( \sum_{j=1}^{p} | \Omega_{jk}^{1} - \Omega_{jk}^{2} | \geq 1 \bigg),
\end{equation*}
where $I(\cdot)$ is an indicator function to determine whether gene $k$ has differential connectivity. The gene $k$ is truly DC if $\eta_{k} = 1$, and is not DC if $\eta_{k} = 0$ for the true gene network. Similarly,  for the covariate dependence structure, the following quantities are obtained
\begin{equation*}
    \eta_{k} = I\bigg( \frac{1}{c} \sum_{c} \sum_{j=1}^{p} | \Omega_{jk}^{1,c} -  \Omega_{jk}^{2, c} | \geq 1 \bigg),
\end{equation*}
where $\Omega^{z, c} \in \mathbf{R}^{p \times p}$ be the adjacency matrix for group $z$ and age category $c=1, 2, 3$. 
Denote that $S$ is the total number of Monte Carlo simulation replicates. Let $q_{ks}$ be adjusted p-value as in following procedure \citep{eBayes2005} of $k^{\text{th}}$ gene at the $s^{\text{th}}$ simulation replicate. $\alpha$ represents the magnitude of error control, and 0.05 was used throughout the simulation. 
\begin{itemize}
\item[-]   Precision is the proportion of genes that are inferred to be significantly DC from the test which have true connection between two comparing groups:
    \begin{equation*}
      \text{Precision} = \frac{ \sum_{k = 1}^{p} \eta_{k} \, I(q_{ks} < \alpha) }{\sum_{k = 1}^{p} I(q_{ks} < \alpha) }.
  \end{equation*}
  \item[-]   Recall is the proportion of genes that have true connection which are correctly inferred to be significantly DC between two comparing groups from the test:
  \begin{equation*}
      \text{Recall} = \frac{ \sum_{k = 1}^{p} \eta_{k} \, I(q_{ks} < \alpha) }{\sum_{k = 1}^{p} \eta_{k} }.
  \end{equation*}
  \item[-]   F1 is calculated based on the harmonic mean of precision and recall obtained from the simulation. A higher F1 score suggests lower false negative and false positive rate:
  \begin{equation*}
      \text{F1} = 2 \cdot \frac{\text{Precision} \cdot \text{Recall}}{\text{Precision} + \text{Recall}}.
  \end{equation*}
\end{itemize}

\subsection{COPDGene Data}
A recent genome-wide association study \citep{COPDgenes} identified 35 new COPD-related genes from the UK Biobank and International COPD Genetics Consortium data. Among these 35 COPD-related genes, 28 genes are available in the data from the COPDGene study for the analysis using PRANA and other DN analysis methods including \textit{dnapath} and DINGO. The 28 COPD-related genes are the following: CITED2, TESK2, COL15A1, AMZ1, RASEF, DDX1, DMWD, MED13L, ZBTB38, CCDC69, EML4, HSPA4, ITGB8, TEPP, TNPO1, ARNTL, DTWD1, ADAMTSL3, RREB1, THRA, SLMAP, DENND2D, STN1, SYN3, ASAP2, IER3, MFHAS1, and VGLL4.

Among 2,561 samples from the initial phenotype data, we have used 406 samples that were provided as the validation set in the analysis of the original study. For the analysis with PRANA, binary current smoking status variable is used as the grouping variable, and smoking pack years, age, gender, race, and FEV1 are included as additional covariates in a multivariable model. The binary current smoking status variable is used as the grouping variable for \textit{dnapath} and DINGO.

\section{Results} 
\subsection{Simulation Study}
More details of the simulation setup are available in the ``Materials'' section above. We select $p = 20, 50, 100$ genes to test our pseudo-value approach in smaller to larger gene networks. For each gene network, five different sample sizes $n = 40, 100, 200, 500, 1000$ are considered and in each setting, 1,000 Monte Carlo replicates. We draw 1,000 random samples first, then take the subsamples from this pool for a simulation with a smaller sample size to reduce computational burden. A random network is generated at each simulation replicate in which a layer of randomness is imposed to account for biological variability of the network structure. For additional details on the generation of simulated RNA-Seq data, see the Materials section. Simulations are repeated to show the performance of our method by altering the effect size from 5$\%$, 10$\%$, to 20$\%$ for simulation scenarios I and II.

Results are compared with the true parent network in order to compute the performance measures described in the ``Performance Measures'' section. In the true network setting, a gene is considered truly DC between groups if it has at least one DC edge connected to other genes. Tables 1 and 2 summarize simulation results in the multivariable setting for scenarios I and II, respectively, when the continuous variable is added as a covariate with the binary group variable in the regression. Table 2 incorporates the effect of covariate when generating random networks, whereas Table 1 does not. 
For both scenarios, results show that the pseudo-value regression method generally yields a high precision and recall across all specifications of network size, sample size, and effect size. The pseudo-value regression method maintains a high precision while having an acceptable recall, especially, when a smaller sample size is considered.

\begin{table}
    \centering
    \setlength{\tabcolsep}{4pt}
\caption{Scenario I simulation results of binary group variable in the multivariable robust regression model (continuous age and binary group) using pseudo-value approach with 1,000 replicates. Random network is generated at each simulation replicate.}{\footnotesize
\begin{tabular}{ll  m{1cm}m{1cm}m{1cm} | m{1cm}m{1cm}m{1cm} | m{1cm}m{1cm}m{1cm} }
 \hline
& & \multicolumn{9}{c}{Effect size} \\ \cline{3-11}
& & \multicolumn{3}{c}{5$\%$} & \multicolumn{3}{c}{10$\%$} & \multicolumn{3}{c}{20$\%$} \\ [0.5ex]
\cline{3-5}\cline{6-8}\cline{9-11}
$p$ & $n$ & Precision & Recall & F1 & Precision & Recall & F1 & Precision & Recall & F1  \\
 \hline
20 & 40 & 0.76 & 0.81 & 0.77 & 0.90 & 0.79 & 0.83 & 0.98 & 0.79 & 0.87 \\
& 100 & 0.75 & 0.91 & 0.82 & 0.90 & 0.91 & 0.90 & 0.98 & 0.91 & 0.94  \\ 
& 200 & 0.74 & 0.94 & 0.82 & 0.89 & 0.94 & 0.91 & 0.97 & 0.94 & 0.96 \\ 
& 500 & 0.73 & 0.97 & 0.83 & 0.88 & 0.97 & 0.92 & 0.97 & 0.97 & 0.97 \\ 
& 1,000 & 0.73 & 0.98 & 0.83 & 0.88 & 0.98 & 0.92 & 0.97 & 0.98 & 0.97 \\ \hline
50 & 40 & 0.95 & 0.65 & 0.77 & 0.98 & 0.65 & 0.78 & 1.00  & 0.64  & 0.77 \\
& 100 & 0.95 & 0.77 & 0.85 & 0.98 & 0.77 & 0.86 & 1.00 & 0.77 & 0.86 \\ 
& 200 & 0.96 & 0.85 & 0.90 & 0.99 & 0.85 & 0.91 & 1.00 & 0.85 & 0.91 \\ 
& 500 & 0.95 & 0.93 & 0.94 & 0.99 & 0.92 & 0.95 & 1.00 & 0.92 & 0.96 \\ 
& 1,000 & 0.95 & 0.96 & 0.95 & 0.98 & 0.96 & 0.97 & 1.00 & 0.96 & 0.98 \\ \hline
100 & 40 & 0.96 & 0.57 & 0.71 & 0.98 & 0.57 & 0.72 & 1.00 & 0.57 & 0.72  \\
& 100 & 0.96 & 0.67 & 0.79 & 0.99 & 0.68 & 0.80 & 1.00 & 0.67 & 0.80 \\
& 200 & 0.96 & 0.74 & 0.83 & 0.99 & 0.74 & 0.84 & 1.00 & 0.74  & 0.85 \\
& 500 & 0.97 & 0.82 & 0.89 & 0.99 & 0.81 & 0.89 & 1.00 & 0.81 & 0.89 \\
& 1,000 & 0.97 & 0.90 & 0.93 & 0.99 & 0.89 & 0.94 & 1.00 & 0.89 & 0.94 \\ \hline
\end{tabular}}{}
\end{table}

\begin{table}
    \centering
    \setlength{\tabcolsep}{4pt}
\caption{Scenario II simulation results of binary group variable in the multivariable robust regression model (continuous age and binary group) using pseudo-value approach with 1,000 replicates. Random network is generated at each simulation replicate.} {\footnotesize
\begin{tabular}{ll  m{1cm}m{1cm}m{1cm} | m{1cm}m{1cm}m{1cm} | m{1cm}m{1cm}m{1cm} }
 \hline
 & & \multicolumn{9}{c}{Effect size} \\ \cline{3-11}
& & \multicolumn{3}{c}{5$\%$} & \multicolumn{3}{c}{10$\%$} & \multicolumn{3}{c}{20$\%$} \\ [0.5ex]
\cline{3-5}\cline{6-8}\cline{9-11}
$p$ & $n$ & Precision & Recall & F1 & Precision & Recall & F1 & Precision & Recall & F1  \\
 \hline
20 & 40 & 0.75 & 0.59 & 0.64 & 0.90 & 0.69 & 0.70 & 0.98 & 0.61 & 0.73 \\
& 100 & 0.78 & 0.70 & 0.71 & 0.91 & 0.72 & 0.79 & 0.98 & 0.74 & 0.83 \\ 
& 200 & 0.78 & 0.81 & 0.78 & 0.91 & 0.83 & 0.86 & 0.98 & 0.85 & 0.91 \\ 
& 500 & 0.76 & 0.90 & 0.82 & 0.90 & 0.92 & 0.91 & 0.98 & 0.94 & 0.95 \\ 
& 1,000 & 0.75 & 0.94 & 0.83 & 0.89 & 0.95 & 0.92 & 0.97 & 0.96 & 0.97 \\ \hline
50 & 40 & 0.95 & 0.56 & 0.70 & 0.98 & 0.56 & 0.71 & 1.00 & 0.57 & 0.72 \\
& 100 & 0.95 & 0.65 & 0.77 & 0.98 & 0.66 & 0.78 & 1.00 & 0.68 & 0.80 \\ 
& 200 & 0.96 & 0.75 & 0.84 & 0.99 & 0.75 & 0.85 & 1.00 & 0.77 & 0.87 \\ 
& 500 & 0.96 & 0.88 & 0.92 & 0.99 & 0.87 & 0.92 & 1.00 & 0.89 & 0.94 \\ 
& 1,000 & 0.96 & 0.93 & 0.94 & 0.99 & 0.93 & 0.96 & 1.00 & 0.94 & 0.97 \\ \hline
100 & 40 & 0.96 & 0.55 & 0.69 & 0.99 & 0.55 & 0.70 & 0.99 & 0.55 & 0.70 \\
& 100 & 0.95 & 0.63 & 0.75 & 0.98 & 0.63 & 0.76 & 0.98 & 0.63 & 0.76 \\
& 200 & 0.96 & 0.68 & 0.79 & 0.99 & 0.68 & 0.81 & 0.99 & 0.68 & 0.80 \\
& 500 & 0.97 & 0.77 & 0.86 & 0.99 & 0.77 & 0.86 & 0.99 & 0.77 & 0.86 \\
& 1,000 & 0.97 & 0.87 & 0.92 & 0.99 & 0.86 & 0.92 & 0.99 & 0.86 & 0.92 \\ \hline
\end{tabular}}{}
\end{table}

Tables 3 and 4 summarize simulation results for scenarios I and II, respectively, when only the binary group variable is included in the model for pseudo-value calculation. Thus, age dependent networks are simulated for Table 4 but not for Table 3. Two competing univariable methods, \textit{dnapath} and DINGO are included for these simulations.

Overall, a similar pattern is observed in the univariable setting; \textit{i.e.}, PRANA consistently reaches a high precision and recall. The performance improves as \textit{n} increases, as to be expected. It is noteworthy that PRANA outperforms \textit{dnapath} in simulation when the sample size is relatively small regardless of the network size. Our method also shows a better recall value and F1-score than \textit{dnapath} with small sample sizes ($\textit{n} = 40, 100$). As DINGO requires substantially large computational time, it was considered for the gene network with smaller sample sizes only. To be more specific, simulations with larger sample sizes ($n = 500, 1000$) are stopped after 20 days for DINGO from the University of Florida Research Computing Linux server, HiPerGator 3.0 with 10CPU cores and 10GB of RAM per node. See Table S1 in Additional File 1 for more details.

\begin{table}
    \centering
    \setlength{\tabcolsep}{4pt}
\caption{Scenario I simulation results of binary group variable in the univariable robust regression model using pseudo-value approach with 1,000 replicates. The network structure does not depend on age covariate. Random network is generated at each simulation replicate. Sample size \textit{n} = (500, 1000) for gene size \textit{p} = 100 were not included for DINGO due to heavy computational time. The best results are highlighted in boldface.}
{\footnotesize
\begin{tabular}{ll  m{1cm}m{1cm}m{1cm} | m{1cm}m{1cm}m{1cm} | m{1cm}m{1cm}m{1cm} }
 \hline
& & \multicolumn{3}{c|}{Precision} & \multicolumn{3}{c|}{Recall} & \multicolumn{3}{c}{F1} \\ [0.5ex]
\cline{3-5}\cline{6-8}\cline{9-11}
$p$ & $n$  & PRANA & \textit{dnapath} & DINGO & PRANA & \textit{dnapath} & DINGO & PRANA & \textit{dnapath} & DINGO \\
 \hline
20 & 40 & 0.90 & \textbf{0.95} & 0.87 & \textbf{0.81} & 0.64 & 0.78 & \textbf{0.84} & 0.76 & 0.82 \\
& 100 & 0.90 & \textbf{0.93} & 0.87 & \textbf{0.90} & 0.88 & 0.79 & 0.84 & \textbf{0.90} & 0.82  \\
& 200 & 0.89 & \textbf{0.91} & 0.87 & \textbf{0.94} & \textbf{0.94} & 0.76 & 0.91 & \textbf{0.92} & 0.81 \\ 
& 500 & 0.88 & \textbf{0.89} & - & 0.97 & \textbf{0.98} & - & 0.92 & \textbf{0.93} & - \\ 
& 1,000 & 0.88 & \textbf{0.89} & - & 0.98 & \textbf{0.99} & - & 0.92 & \textbf{0.93} & -\\ \hline
50 & 40 & 0.98 & \textbf{1.00} & 0.99 & 0.65 & 0.39 & \textbf{0.70} & 0.78 & 0.56 & \textbf{0.81} \\
& 100 & 0.98 & \textbf{1.00} & 0.98 & 0.77 & 0.61 & \textbf{0.84} & 0.86 & 0.75 & \textbf{0.90} \\ 
& 200 & 0.99 & \textbf{1.00} & 0.98  & \textbf{0.85} & \textbf{0.85} & \textbf{0.85} & \textbf{0.91} & \textbf{0.91} & \textbf{0.91}  \\ 
& 500 & \textbf{0.99} & \textbf{0.99} & - & 0.92 & \textbf{0.95} & - & 0.95 & \textbf{0.97} &  - \\
& 1,000 & 0.98 & \textbf{0.99} & - & 0.96 & \textbf{0.98} & - & 0.97 & \textbf{0.98} & - \\ \hline
100 & 40 & 0.98 & \textbf{1.00} & 0.98 & 0.57 & 0.27 & \textbf{0.69} & 0.72 & 0.42 & \textbf{0.85} \\
& 100 & 0.98 & \textbf{1.00} & 0.99 & 0.68 & 0.32 & \textbf{0.75}  & 0.80 & 0.48 & \textbf{0.85} \\ 
& 200 & 0.99 & \textbf{1.00} & 0.98 & 0.74 & 0.62 & \textbf{0.82} & 0.84 & 0.77 & \textbf{0.89} \\ 
& 500 & 0.99 & \textbf{1.00} & - & 0.81 & \textbf{0.88} & - & 0.89 & \textbf{0.93} & - \\
& 1,000 & \textbf{0.99} & \textbf{0.99} & - & 0.89 &  \textbf{0.94} & - & 0.94 & \textbf{0.97} & - \\ \hline
\end{tabular}}{}
\end{table}

\begin{table}
    \centering
    \setlength{\tabcolsep}{4pt}
\caption{Scenario II simulation results of binary group variable in the univariable robust regression model using pseudo-value approach with 1,000 replicates. The network structure depends on age covariate. Random network is generated at each simulation replicate. Sample size \textit{n} = (500, 1000) or gene size \textit{p} = 100 were not included for DINGO due to heavy computational time. The best results are highlighted in boldface.}
{\footnotesize
\begin{tabular}{ll  m{1cm}m{1cm}m{1cm} | m{1cm}m{1cm}m{1cm} | m{1cm}m{1cm}m{1cm} }
 \hline
& & \multicolumn{3}{c|}{Precision} & \multicolumn{3}{c|}{Recall} & \multicolumn{3}{c}{F1} \\ [0.5ex]
\cline{3-5}\cline{6-8}\cline{9-11}
$p$ & $n$  & PRANA & \textit{dnapath} & DINGO & PRANA & \textit{dnapath} & DINGO & PRANA & \textit{dnapath} & DINGO \\
 \hline
20 & 40 & 0.90 & \textbf{0.97} & 0.89 & 0.59 & 0.38 & \textbf{0.63} & 0.69 & 0.53 & \textbf{0.72} \\
& 100 & 0.91 & \textbf{0.97} & 0.88 & 0.72 & 0.66 & \textbf{0.75} & 0.79 & 0.77 & \textbf{0.80} \\
& 200 & 0.91 & \textbf{0.96} & 0.89 & \textbf{0.83} & \textbf{0.83} & 0.73 & 0.86 & \textbf{0.88} & 0.79 \\ 
& 500 & 0.90 & \textbf{0.93} & - & 0.93 & \textbf{0.94} & - & 0.91 & \textbf{0.93} & - \\ 
& 1,000 & 0.89 & \textbf{0.91} & - & 0.95 & \textbf{0.97} & - & 0.92 & \textbf{0.94} & - \\ \hline
50 & 40 & 0.98 & \textbf{1.00} & 0.99 & 0.56 & 0.28 & \textbf{0.67} & 0.71 & 0.43 & \textbf{0.80} \\
& 100 & 0.98 & \textbf{1.00} & 0.98 & 0.65 & 0.45 & \textbf{0.71} & 0.78 & 0.61 & \textbf{0.82} \\ 
& 200 & 0.99 & \textbf{1.00} & 0.98 & 0.75 & 0.68 & \textbf{0.80} & 0.85 & 0.80 & \textbf{0.88} \\ 
& 500 & \textbf{0.99} & \textbf{0.99} & - & 0.87 & \textbf{0.91} & - & 0.93 & \textbf{0.95} & - \\
& 1,000 & \textbf{0.99} & \textbf{0.99} & - & 0.94 & \textbf{0.97} & - & 0.96 & \textbf{0.98} & - \\ \hline
100 & 40 & 0.99 & \textbf{1.00} & 0.55 & 0.55 & 0.21 & \textbf{0.77} & \textbf{0.70} & 0.34 & 0.63 \\
& 100 & 0.98 & \textbf{1.00} & 0.55 & 0.63 & 0.27 & \textbf{0.75} & \textbf{0.76} & 0.42 & 0.63 \\ 
& 200 & 0.99 & \textbf{1.00} & 0.55 & 0.68 & 0.46 & \textbf{0.77} & \textbf{0.80} & 0.62 & 0.63 \\ 
& 500 & 0.99 & \textbf{1.00} & - & 0.77 & \textbf{0.82} & - & 0.86 & \textbf{0.90} & - \\
& 1,000 & 0.99 & \textbf{1.00} & - & 0.86 & \textbf{0.92} & - & 0.92 & \textbf{0.96} & - \\ \hline
\end{tabular}}{}
\end{table}

Table 5 presents results of scenario III, where age acts as a confounder. That is, an observed difference in connectivity may be due to a difference in the distribution of age in the two groups. Higher precision values from the multivariable pseudo-value regression indicate that PRANA correctly identifies the DC genes, compared with \textit{dnapath} and DINGO, neither of which accounts for the effects of age. By and large, PRANA has higher precision than DINGO and higher recall than \textit{dnapath}.

\begin{table}
    \centering
    \setlength{\tabcolsep}{4pt}
\caption{Scenario III simulation results of binary group variable in the multivariable and univariable robust regression model using pseudo-value approach with 1,000 replicates. The network structure depends on age covariate by unequal sampling proportion depending on age categories. Random network is generated at each simulation replicate. Sample size \textit{n} = (500, 1000) or gene size \textit{p} = 100 were not included for DINGO due to heavy computational costs. The best results are highlighted in boldface.}
{\footnotesize
\begin{tabular}{m{.3cm}m{.7cm}  m{0.9cm}m{0.9cm}m{0.9cm}m{0.9cm} | m{0.9cm}m{0.9cm}m{0.9cm}m{0.9cm} | m{0.9cm}m{0.9cm}m{0.9cm}m{0.9cm} }
 \hline
& & \multicolumn{4}{c|}{Precision} & \multicolumn{4}{c|}{Recall} & \multicolumn{4}{c}{F1} \\ [0.5ex]
\cline{3-6}\cline{7-10}\cline{11-14}
$p$ & $n$ & PRANA (Mult) & PRANA (Univ) & \textit{dnapath} & DINGO & PRANA (Mult) & PRANA (Univ) & \textit{dnapath} & DINGO & PRANA (Mult) & PRANA (Univ)& \textit{dnapath} & DINGO \\
 \hline
20 & 40 & \textbf{0.67} & 0.60 & 0.64 & 0.58 & 0.57 & 0.74 & 0.50 & \textbf{0.75} & 0.59 & \textbf{0.65} & 0.55 & 0.64 \\
& 100 & \textbf{0.67} & 0.58 & 0.61 & 0.58 & 0.65 & \textbf{0.85} & 0.73 & 0.79 & 0.64 & \textbf{0.68} & 0.65 & 0.66 \\
& 200 & \textbf{0.66} & 0.58 & 0.59 & 0.58 & 0.76 & \textbf{0.91} & 0.85 & 0.80 & 0.69 & \textbf{0.70} & 0.69 & 0.67 \\ 
& 500 & \textbf{0.64} & 0.57 & 0.58 & - & 0.87 & \textbf{0.95} & \textbf{0.95} & - & \textbf{0.73} & 0.71 & 0.71 & - \\ 
& 1,000 & \textbf{0.64} & 0.57 & 0.58 & - & 0.92 & \textbf{0.97} & \textbf{0.97} & - & \textbf{0.75} & 0.71 & 0.72 & - \\ \hline
50 & 40 & \textbf{0.57} & 0.50 & 0.54 & 0.49 & 0.47 & 0.62 & 0.33 & \textbf{0.67} & 0.50 & 0.54 & 0.40 & \textbf{0.55} \\
& 100 & \textbf{0.58} &  0.49 & 0.52 & 0.49 & 0.50 & 0.71 & 0.51 & \textbf{0.79} & 0.52 & 0.58 & 0.51 & \textbf{0.60} \\ 
& 200 & \textbf{0.58} & 0.49 & 0.51 & 0.49 & 0.52 & 0.76 & 0.60 & \textbf{0.83} & 0.53 & 0.59 & 0.54 & \textbf{0.61} \\ 
& 500 & \textbf{0.55} & 0.48 & 0.48 & - & 0.67 & \textbf{0.88} & 0.83 & - & 0.59 & \textbf{0.61} & 0.60 & - \\
& 1,000 & \textbf{0.54} & 0.48 & 0.48 & - & 0.81 & \textbf{0.93} & 0.91 & - & \textbf{0.64} & 0.62 & 0.62 & -\\ \hline
100 & 40 & \textbf{0.57} & 0.49 & 0.53 & 0.48 & 0.44 & 0.57 & 0.22 & \textbf{0.75} & 0.49 & 0.52 & 0.30 & \textbf{0.58} \\
& 100 & \textbf{0.58} & 0.49 & 0.53 & 0.48 & 0.47 & 0.64 & 0.35 & \textbf{0.76} & 0.51 & 0.55 & 0.41 & \textbf{0.58} \\ 
& 200 & \textbf{0.58} & 0.49 & 0.53 & 0.48 & 0.46 & 0.65 & 0.40 & \textbf{0.80} & 0.50 & 0.55 & 0.45 & \textbf{0.60} \\ 
& 500 & \textbf{0.56} & 0.47 & 0.49 & - & 0.44 & \textbf{0.74} & 0.60 & - & 0.48 & \textbf{0.57} & 0.53 & - \\
& 1,000 & \textbf{0.53} & 0.47 & 0.47 & - & 0.63 & \textbf{0.85} & 0.82 & - & 0.57 & \textbf{0.60} & 0.59 & - \\ \hline
\end{tabular}}{}
\end{table}

\begin{table}
\renewcommand\thetable{S1} 
    \centering
    \setlength{\tabcolsep}{6pt}
\caption{Comparison of computational time of PRANA with that of \textit{dnapath} and DINGO. Scenario I is depicted for the illustrative purposes. Random network is generated at each simulation replicate. Large sample sizes \textit{n} = (500, 1000) are stopped after 20 days for DINGO from the high-performance Linux cluster using machines with 10CPU cores and 10GB of RAM per node.}
{\small
\begin{tabular}{ll  m{2cm}m{2cm}m{2cm}  }
 \hline
& & \multicolumn{3}{c}{Time (in hours)}  \\ [0.5ex]
$p$ & $n$  & PRANA & \textit{dnapath} & DINGO  \\
 \hline
20 & 40 & 0.12 & 0.13 & 29.25 \\
& 100 & 0.22 & 0.2 & 86.58 \\
& 200 & 0.23 & 0.17 & 206.18 \\
& 500 & 0.42 & 0.27 & $>$ 480 \\
& 1,000 & 0.68 & 0.45 &  $>$ 480 \\ \hline
50 & 40 & 0.2 & 0.22 & 26.57 \\
& 100 & 0.35 & 0.32 & 31.48  \\
& 200 & 0.67 & 0.45 & 137.58 \\
& 500 & 1.52  & 0.98 & $>$ 480 \\
& 1,000 & 2.48 & 1.08 & $>$ 480 \\ \hline
100 & 40 & 0.63 & 0.52 & 68.73 \\
& 100 & 0.82 & 0.7 & 61.78 \\
& 200 & 1.1 & 1.23 & 147.57  \\
& 500 & 3.58 & 2.37 & $>$ 480 \\
& 1,000 & 4.68 & 3.57 & $>$ 480 \\ \hline
\end{tabular}}{}
\end{table}

\subsection{Analysis of COPDGene Data}
23 out of 28 COPD-related genes are predicted to be DC between current and non-current smokers with PRANA while accounting for smoking pack years, age, gender, race, and FEV1. A complete list of DC genes found from the pseudo-value approach are CITED2, TESK2, AMZ1, DDX1, DMWD, MED13L, ZBTB38, EML4, HSPA4, ITGB8, TEPP, TNPO1, ARNTL, DTWD1, ADAMTSL3, THRA, SLMAP, DENND2D, STN1, SYN3, ASAP2, IER3, and MFHAS1. 

We compared results of PRANA with \textit{dnapath} \citep{dnapath} and DINGO \citep{DINGO}. With DINGO, a total of 19 out of 28 COPD-related genes were selected as DC genes between current and non-current smokers. A complete list of DC genes found in DINGO are the following: ARNTL, DDX1, HSPA4, ITGB8, SLMAP, SYN3, ASAP2, IER3, MFHAS1, VGLL4, CITED2, TESK2, CCDC69, EML4, ADAMTSL3, DENND2D, AMZ1, RASEF, and ZBTB38. Lastly, 3 genes were found DC between current smoking groups with \textit{dnapath}, namely DTWD1, EML4, and TEPP.

Of the 23 DC genes from PRANA, 5 are found exclusive to PRANA (DMWD, MED13L, TNPO1, THRA, and STN1). Notably, DMWD is linked to myotonic dystrophy, a rare genetic muscular disorder \citep{DMWD}. Thyroid hormone receptor alpha (THRA) is related to congenital hypothyroidism \citep{THRA}. These findings about additional genes will facilitate harnessing of the possible mechanisms at work in COPD exacerbation. 

Heat shock protein family A (Hsp70) member 4 (HSPA4) is associated with gastric ulcer \citep{HSPA4}. Multifunctional ROCO family signaling regulator 1 (MFHAS1) is linked to soft tissue tumor and cell cycle \citep{GeneCards}. HSPA4 and MFHAS1 are DC genes identified in both PRANA and DINGO. Echinoderm microtubule-associated protein-like 4 (EML4) is found in all three methods. It has been studied for its association with lung cancer \citep{GeneCards, EML4_2}. A Venn diagram is provided to show the overlap between and among three methods (Figure 4). In addition, a diagram is included to summarize the findings of this application study (Figure 5).

\begin{figure}
 \centerline{\includegraphics[scale=0.6]{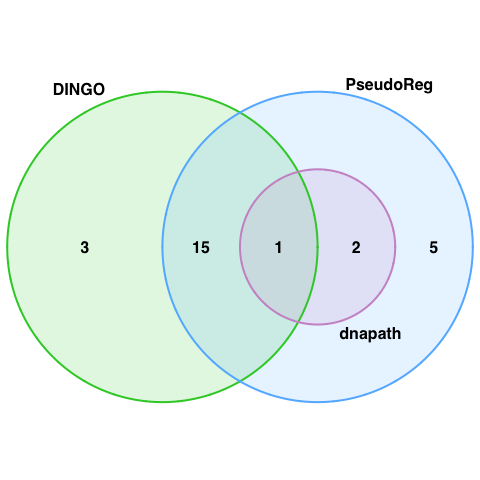}}
  \caption{A Venn diagram displaying the number of overlapping DC genes between and among univariable analysis such as DINGO and \textit{dnapath} versus multivariable robust regression with pseudo-value approach using COPDGene study data from GEO database.}
\end{figure}

\begin{figure}
 \centerline{\includegraphics[scale=0.9]{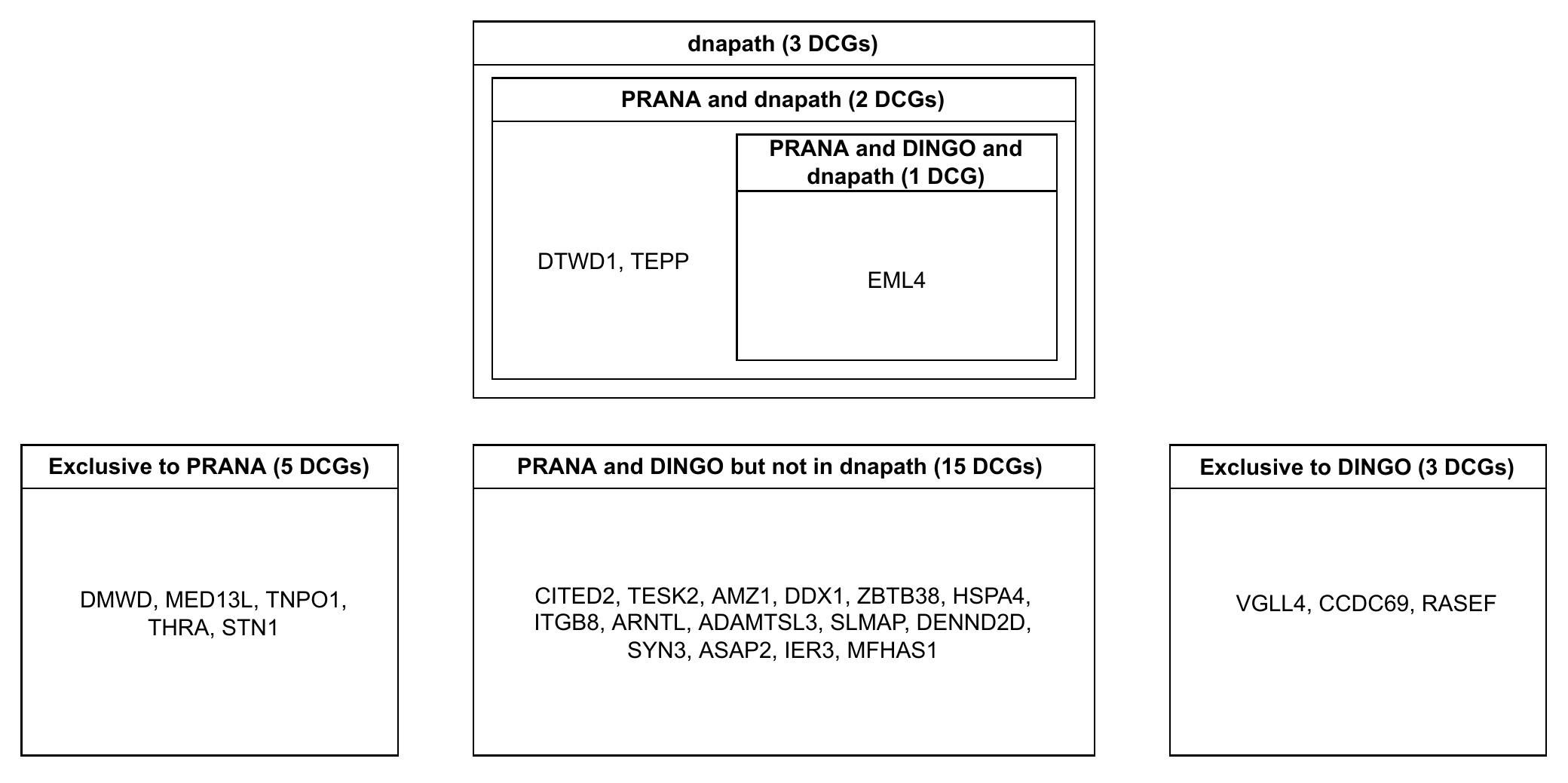}}
\caption{A diagram summarizing results using each methods analyzing the COPDGene study data from GEO database. A full list of DCGs (differentially connected genes) are provided in each box.}
\end{figure}

\section{Discussion}
\label{s:discuss}
Simulations and real-data analysis have elucidated that PRANA is superior to existing alternatives and a practical tool, which includes covariates in the model. To the best of our knowledge, this is the first attempt to develop a regression modeling in DN analysis. Our working objective is to propose a statistical method that determines whether a gene is significantly DC between groups with the covariate included in the model. In this paper, we have shown through simulations that PRANA reaches a consistently high degree of precision and recall to identify DC genes with varying simulation parameters such as network size, sample size, and effect size. We also analyzed a COPD-related gene expression data from the GEO database. When comparing results from our method to \textit{dnapath} and DINGO, five COPD-related genes are additionally found DC between current versus non-current smokers: DMWD, MED13L, TNPO1, THRA, and STN1.

There are a number of limitations to be highlighted in this study. We have used the absolute value of the differences between the two adjacency matrices as a proxy to determine the true DC genes. Certainly, this is a practical way to detect differences in the number of edges for each genes in a network. The comparison of maximum values between adjacency matrices was also considered. However, we concluded that they are more useful describing the global characteristic of a network, which deviates from our objective, namely, a gene-specific characteristic of a network.

Another limitation is the inability to perturb simulated networks in a continuous way. Right now, we have discretized the effect of a covariate into three groups. Perhaps, there are other models where a truly continuous covariate could be incorporated.

Lastly, the Pearson correlation, partial correlation, and degree-weighted LASSO were also examined as alternatives to the ARACNE as a measure of association or connectedness, albeit not reported in the paper, due to relatively poor performance and heavy computational costs. It remains an interesting task for future studies to extend our work to other measures of association of a network which better assess different structural changes in the network. We conclude by foregrounding the future direction of the pseudo-value regression approach for the DN analysis, which are potentially extensible to other data types, such as the microbiome data.

\section{Conclusion}
The adjustment of covariate is an important step in differential network analysis. In this paper, we presented PRANA, a novel pseudo-value regression approach for the DN analysis, which can incorporate additional clinical covariates in the model. This is a direct regression modeling, and it is therefore computationally amenable for the most users.


\backmatter





%
\bibliographystyle{biom}
\bibliography{mybib.bib}


\appendix




\label{lastpage}

\end{document}